\documentclass[12pt,twocolumn]{iopart}

\usepackage{graphicx}
\usepackage{amsmath}
\usepackage{xcolor}

\usepackage{iopams}  

\newcommand{\rhob}{\bar\rho}
\newcommand{\hex}{h_{\rm ex}}
\newcommand{\hup}{\overline{h}_{\rm ex}}

\newcommand{\tilv}{\tilde{v}}
\newcommand{\trho}{\tilde{\rho}}
\newcommand{\veck}{{\bf k}}
\newcommand{\vecr}{{\bf r}}

\begin{document}

\title{Fluids with power-law repulsion: Hyperuniformity and energy fluctuations}

\author{H Diamant$^1$, E C O\v{g}uz$^{2,3*}$}
\address{$^1$ School of Chemistry and the Center for Physics and Chemistry of Living Systems, Tel Aviv University, Tel Aviv 6997801, Israel}
\address{$^2$ Key Laboratory of Soft Matter Physics, Institute of Physics, Chinese Academy of Sciences, Beijing 100190, China}
\address{$^3$ Songshan Lake Materials Laboratory, Dongguan, Guangdong 523808, China}

\ead{ecoguz@iphy.ac.cn}

\vspace{10pt}
%\begin{indented}
%\item[]December 2023
%\end{indented}

\begin{abstract}

We revisit the equilibrium statistical mechanics of a classical fluid of point-like particles with repulsive power-law pair interactions, focusing on density and energy fluctuations at finite temperature. Such long-range interactions, decaying with inter-particle distance $r$ as $1/r^s$ in $d$ dimensions, are known to fall into two qualitatively different categories. For $s<d$ (``strongly" long-range interactions) there are screening of correlations and suppression of large-wavelength density fluctuations (hyperuniformity). These effects eliminate density modes with arbitrarily large energy. For $s>d$ (``weakly" long-range interactions) screening and hyperuniformity do not occur. Using scaling arguments, variational analysis, and Monte Carlo simulations, we find another qualitative distinction. For $s\geq d/2$ the strong repulsion at short distances leads to enhanced small-wavelength density fluctuations, decorrelating particle positions. This prevents indefinitely negative entropy and large energy fluctuations. The distinct behaviors for $s\geq d/2$ and $s<d/2$ give rise to qualitatively different dependencies of the entropy, heat capacity, and energy fluctuations on temperature and density. We investigate the effect of introducing an upper cutoff distance in the pair-potential. The effect of the cutoff on energy fluctuations is strong for $s<d/2$ and negligible for $s\geq  d/2$. 
%cutoff distance on the suppression of the large-wavelength density fluctuations for $s<d$, and find that the cutoff  originates unusual oscillatory behavior of the structure factor and the pair-correlation function.}

\end{abstract}

\vspace{2pc}

%\maketitle

\section{Introduction}

Various physical systems contain long-range interactions which decay algebraically with the distance between particles. These include, for example, coulombic and dipolar fluids, plasmas, topological defects in solids and liquid crystals, inclusions in elastic media and membranes, and vortices in superfluids. The statistical physics of such systems has been extensively studied; see reviews with comprehensive literature in refs~\cite{CampaPhysRep2009,BouchetPhysA2010,LewinJMathPhys2022}. The long-range interactions lead to various distinctive behaviors and anomalies, such as screening of correlations~\cite{BrydgesCommMathPhys1980,Alastuey1985,BrydgesJSP1999}, super-extensive thermodynamic potentials~\cite{Anteneodo1998} and sub-extensive fluctuations~\cite{MartinJSP1980,LebowitzPRA1983,JancoviciJSP1993}, repulsion-stabilized (Wigner) crystallization~\cite{ThomasPRL1994,AboShaeerScience2001}, and strong surface effects that raise issues concerning the thermodynamic limit~\cite{Lieb1972,Gregg1989} and the equivalence of statistical ensembles~\cite{Barre2001,Bouchet2005,Ellis2004,Leyvraz2002}.

One of the anomalies occurring for repulsive power-law potentials is the suppression of long-wavelength density fluctuations, or hyperuniformity \cite{TorquatoPRE2003,TorquatoPhysRep2018,LewinJMathPhys2022}, which is directly tied to the sub-extensive fluctuations of particle number~\cite{MartinJSP1980,LebowitzPRA1983,JancoviciJSP1993}. If the pair-potential decays as $1/r^s$ with $s<d$ ($d$ being the number of dimensions), then the structure factor of the corresponding fluid, $S(\veck)$, which normally reaches a constant at small wavevectors $k$, instead decays to zero  as $S(\veck)\sim k^\alpha$, $\alpha=d-s$. The wavevector below which this decay takes place defines a screening length, beyond which the density fluctuations are suppressed. In section~\ref{sec:scaling} we describe the emergence of hyperuniformity and the screening length based on intuitive scaling arguments.

A large body of rigorous theoretical work has accumulated over the years concerning systems with power-law interactions~\cite{CampaPhysRep2009,BouchetPhysA2010,LewinJMathPhys2022,Serfaty2014,BausPhysRep1980,Forrester1998,HooverJCP1971,GruberJSP1980}. These studies address, for example, the stability of crystalline structures~\cite{PollockPRA1973,Travesset2014}, fluid-solid transitions~\cite{AgrawalPRL1995}, the screening phenomenon~\cite{BrydgesCommMathPhys1980,BrydgesJSP1999}, sum rules associated with density correlations~\cite{BlumPRL1982,MartinRMP1988}, and particle-number fluctuations~\cite{MartinJSP1980,LebowitzPRA1983,JancoviciJSP1993}. The present work does not continue this line of studies. Our goal is to turn the spotlight toward a different distinctive behavior of these systems, which to our knowledge has not been recognized. The indefinitely strong repulsion at {\em short} inter-particle distances, for $s\geq d/2$, leads to anomalous suppression of small-wavelength (large-$k$) density 
correlations. Such small-scale features are usually associated with non-universal molecular details and assumed not to affect the qualitative thermodynamics for densities that are not too high. However, we find that in the case of a power-law repulsion with $s\geq d/2$ this decorrelation \cite{SalDecorr2006}
%``small-scale hyperuniformity"
 qualitatively changes the fluid's thermodynamics, namely, how its entropy, heat capacity, and energy fluctuations scale with temperature and density, even at high temperature or low density. The relevant small length scale, akin to the Bjerrum length in coulombic fluids~\cite{LevinRepProgPhys2002}, is independent of the molecular size and typically larger.

In section \ref{sec:system} we define the problem and its parameters and present its characteristic length scales. In section \ref{sec:scaling} we obtain some of the qualitative results that follow, based on a simple scaling analysis. Section \ref{sec:variational} provides a variational analysis which yields the high-temperature structure factor and entropy. In section \ref{sec:fluctuations} we show that for $s\geq d/2$ the results of section \ref{sec:variational} yield a divergent entropy. We correct this failure using an intrinsic upper cutoff for $k$, obtaining modified scaling laws for the entropy and energy fluctuations. Section~\ref{sec:simulations} presents Monte Carlo simulations, which confirm the analytical predictions and explore the effect of cutting off the power-law potential at a finite, large inter-particle distance. Finally, in section~\ref{sec:discussion}, we discuss the results and their experimental relevance.

\section{The system}
\label{sec:system}

We consider an equilibrium fluid of $N$ point-like particles in $d$-dimensional volume $V$ and at finite temperature (in energy units) $T$. The mean density is $\rhob=N/V$. The particles interact via the repulsive pair-potential
\begin{equation}
    v(\vecr) =  \frac{v_0}{r^s},\ \ \ v_0>0,\ \ \ 
    s \geq 0.
\label{vr}
\end{equation}
This model is sometimes referred to as the repulsive Riesz gas \cite{LewinJMathPhys2022}. For a 3D coulombic fluid $s=1$. 
The $d$-dimensional Fourier transform of the pair-potential (properly regularized) is
\begin{equation}
    \tilv(\veck) = \frac{b v_0}{k^{d-s}},
\label{vk}
\end{equation}
where $b$ is a known numerical prefactor which depends on $d$ and $s$. For a 3D coulombic fluid $d-s=2$ and $b=4\pi$.

The model has two intrinsic lengths, the mean inter-particle distance, $\rhob^{-1/d}$, and the distance below which the pair-interaction energy is larger than $T$, $\ell=(v_0/T)^{1/s}$. For a 3D coulombic fluid $\ell=v_0/T$ is known as the Bjerrum length~\cite{LevinRepProgPhys2002}. The two lengths define high- and low-temperature (equivalently, dilute and concentrated) regimes, $\ell\rhob^{1/d}\ll 1$ and $\ell\rhob^{1/d}\gg 1$, respectively. Out of the two lengths emerges a screening length, $\lambda \sim \ell^{-s/(d-s)} \rhob^{-1/(d-s)}$, as discussed below. For a 3D coulombic fluid $\lambda\sim (\ell\rhob)^{-1/2}$ is the Debye screening length \cite{LevinRepProgPhys2002}.

We are interested in details of the equilibrium two-point correlation function of density fluctuations, $\langle\rho({\bf 0})\rho(\vecr)\rangle$, or its Fourier transform, the structure factor,
\begin{equation}
    S(\veck) = \rhob^{\,-1} \langle |\trho(\veck)|^2 \rangle,
\end{equation}
and its effect on the fluid's thermodynamics.

\section{Scaling analysis}
\label{sec:scaling}

The fact that the potential (\ref{vr}) has no characteristic length scale allows for several key results to be obtained from simple scaling arguments.

First, in the case of normal density fluctuations, $S(\veck)$ tends to a positive constant at small $k$ \cite{HansenBook}. If this were the case for the potential (\ref{vr}) with $s<d$, the  energy of small-$k$ density modes, $\tilv(\veck)S(\veck)$, would diverge in the limit $k\rightarrow 0$ as $k^{-(d-s)}$. Avoiding it requires hyperuniformity, i.e., the structure factor must tend to zero at small $k$ as $S(\veck) \sim k^\alpha$, $\alpha\geq d-s$.

To see in more detail how this is brought about, consider a region of size $R$ in the fluid. The mean number of particles in the region is $\langle n\rangle \sim \rhob R^d$. If the number fluctuations inside the region are normal, the variance of the number of particles is $\langle (\delta n)^2\rangle \sim \langle n\rangle \sim \rhob R^d$. The number fluctuation $\delta n$ entails the energy $\langle |\delta u| \rangle \sim \langle (\delta n)^2\rangle\, v(R) \sim \rhob v_0 R^{d-s}$. This energy is provided by the thermal bath,  $\langle |\delta u| \rangle \sim T$, leading to
\begin{equation}
    R = \lambda \sim \left( \frac{T}{\rhob v_0} \right)^{1/(d-s)}.
\label{lambda}
\end{equation}
If $s<d$, normal density fluctuations over distances larger than $\lambda$ require energy larger than $T$ and are suppressed. Thus, if the repulsive potential is sufficiently long-ranged, $s<d$, screening occurs\,---\,normal density fluctuations are confined to within a screening length $\lambda$, which gets smaller with increasing density or decreasing temperature. 
For $s>d$ equation~\eqref{lambda} gives the more common case of a correlation length $\lambda$ getting larger with increasing $\rhob$ or decreasing $T$. In this case normal density fluctuations with energy less than $T$ occur over large length scales $R>\lambda$, while for $R<\lambda$ fluctuations cost more than $T$ due to the (short-range) interactions.
%For $s=d$ the energy of a density fluctuation is $\langle |\delta u|\rangle \sim \rhob v_0$, independent of length scale.

For $s<d$, beyond the screening length, density fluctuations are anomalously small; the fluid is hyperuniform \cite{TorquatoPhysRep2018}. Let us assume that the number fluctuation for $R>\lambda$ is $\langle(\delta n)^2\rangle \sim \rhob^{\gamma/d}\, R^{\gamma}$ with $\gamma<d$. (The power of $\rhob$ is dictated by units.) The number fluctuation cannot grow with $R$ more slowly than the region's surface $R^{d-1}$ \cite{BeckActaMath1987}, i.e., the fluid cannot be more uniform than a crystal. Hence, $\gamma\geq d-1$. The energy  associated with this number fluctuation is now $\langle |\delta u| \rangle \sim \langle(\delta n)^2\rangle\, v(R) \sim v_0\, \rhob^{\gamma/d}\, R^{\gamma-s}$. Since it scales with $R$ with an exponent $\gamma-s < d-s$, this energy is smaller than the one of normal fluctuations. It cannot decrease with $R$ because this would favor strong large-wavelength fluctuations and lead to instability. Hence, $\gamma \geq \max(d-1, s)$. The number fluctuation will have the smallest energy cost allowed for $\gamma=s$. Thus $\gamma=s$ if $s \geq d-1$, and $\gamma=d-1$ if $s < d-1$.
At the same time, the number-fluctuation exponent $\gamma$ is mathematically tied to the structure-factor exponent $\alpha$ describing the approach of $S(\veck)$ to zero at small $k$, $S(\veck)\sim k^\alpha$. The relation is $\gamma=d-1$ if $\alpha>1$ and $\gamma=d-\alpha$ if $\alpha < 1$ \cite{TorquatoPhysRep2018}. (These two cases are referred to, respectively, as class-I and class-III hyperuniformity; the marginal case $\alpha=1$ is termed class-II hyperuniformity \cite{TorquatoPhysRep2018}. These relations have also been generalized to include some aperiodic systems  \cite{Oguz2017PRB,Oguz2019Acta}.)  From the last two results it follows that $\alpha=d-s$.

To sum up the results of the scaling analysis: (a) for $s<d$ there are screening and hyperuniformity, (b) in this case we have at large $R$ or small $k$ the following hyperuniformity features:
\begin{eqnarray}
    &\langle (\delta n)^2 \rangle \sim R^{\gamma},
    \ \ \ \ & \gamma = \left\{ 
    \begin{array}{ll}
      s, & s \geq d-1\\
      d-1, & s < d-1
    \end{array} \right. \\
    &S(\veck) \sim k^\alpha, & \alpha = d-s,
    \label{Sqscaling}
\end{eqnarray}
and (c) these features take place beyond the screening length $\lambda$ given by equation~(\ref{lambda}).

\section{Structure from variational principle}
\label{sec:variational}

A given structure factor $S(\veck)$ sets an upper bound for the entropy \cite{Ariel2020,Sorkin2023},
\begin{equation}
    \hex[S(\veck)] \,\leq\, \hup[S(\veck)] = \frac{1}{2(2\pi)^d \rhob}
    \int d\veck \left[ \ln S(\veck) - S(\veck) + 1 \right].
\label{hex}
\end{equation}
In equation~(\ref{hex}) $\hex$ is the excess entropy per particle over that of uncorrelated particles ($S(\veck)= 1$). The interaction energy per particle can be written in terms of the structure factor as
\begin{equation}
    u[S(\veck)] = \frac{1}{2(2\pi)^d} \int d\veck\, \tilv(\veck)
    \left[ S(\veck) - 1 \right],
\label{u}
\end{equation}
where we have subtracted a constant $\sim\int d\veck\, \tilv(\veck)$ which diverges for $s>0$. While Eq. (\ref{u}) has been utilized to construct ground states that are disordered and hyperuniform \cite{SalPRX2015}, we focus here on the high-temperature limit.
Equations (\ref{hex}) and (\ref{u}) give a lower bound for the excess free energy per particle,
\begin{equation}
    f_{\rm ex}[S(\veck)] = u - T \hex \,\geq\, \underline{f}_{\rm\, ex}[S(\veck)] = u - T \hup. 
\end{equation}
Setting $\delta \underline{f}_{\rm \, ex}/\delta S(\veck) = 0$ gives
\begin{equation}
    S(\veck) = \frac{1} {1 + (\rhob/T) \tilv(\veck)}.
\label{Sk}
\end{equation}
Since $\delta^2 \underline{f}_{\rm \, ex}/\delta S(\veck) \delta S(\veck') \sim [1/S(\veck)^2] \delta(\veck-\veck') > 0$, the structure factor (\ref{Sk}) is a minimizer of the lower bound $\underline{f}_{\rm \, ex}$.

Equation (\ref{Sk}) coincides with a known result for the structure factor of a fluid in the high-temperature regime \cite{HansenBook}. Thus the bound becomes tight in this limit. Specializing equation~(\ref{Sk}) to the potential of equation~(\ref{vk}), we get
\begin{equation}
    S(\veck) = \frac{k^\alpha} {k^\alpha + \lambda^{-\alpha}},\ \ \ \
    \lambda = \left( \frac{T}{b v_0 \rhob} \right)^{1/\alpha},\ \ \ \ \alpha=d-s.
\label{Sqvariational}
\end{equation}
For a 3D coulombic fluid this gives the known structure factor with $\alpha=2$ and the Debye screening length, $\lambda=[T/(4\pi v_0\rhob)]^{1/2}$ \cite{HansenBook}. 

The results above agree with the scaling analysis, equations~(\ref{lambda}) and (\ref{Sqscaling}). Moreover, the limiting behavior of equation~(\ref{Sqvariational}), $S(k\ll\lambda^{-1})\sim k^\alpha$, is demanded by exact sum rules at all temperatures \cite{BlumPRL1982,MartinRMP1988}. Thus, although the equality $\hex=\hup$ and the structure factor (\ref{Sqvariational}) are strictly valid only in the high-temperature limit, we expect them to be qualitatively applicable also away from this limit. This will be confirmed by simulations in section~\ref{sec:simulations} for $s<d/2$. For $s\geq d/2$, the analyses given above fail, as discussed in the next section.

\section{Entropy and energy fluctuations}
\label{sec:fluctuations}

Substituting $S(\veck)$ of equation~(\ref{Sqvariational}) back in the integral of equation~(\ref{hex}) gives an upper bound for the entropy, which depends only on $\rhob\lambda^d$ and $s/d$. For $s/d<1/2$, the result is
\begin{equation}
\hspace{-1cm}
    \hup = \frac{{\cal S}_d} {2(2\pi)^d d}\, \frac{1} {\rhob \lambda^d}\, \psi(s/d),\ \ \  \psi(x)=\frac{\pi x}{1-x}\, \frac{1}{\sin[\pi/(1-x)]},\ \ \  s/d<1/2,
\label{divergence}
\end{equation}
where ${\cal S}_d$ is the surface of the unit sphere in $d$ dimensions. For a 3D coulombic fluid ${\cal S}_d=4\pi$ and $\psi(1/3)=-1/6$, yielding $\hup=-(24\pi\rhob\lambda^3)^{-1}$, which coincides with the entropy obtained from the Debye-H\"uckel theory. Thus the bound is equal indeed to the actual entropy in the high-temperature limit.

\begin{figure}
\centerline{
 \includegraphics[width=0.7\textwidth]{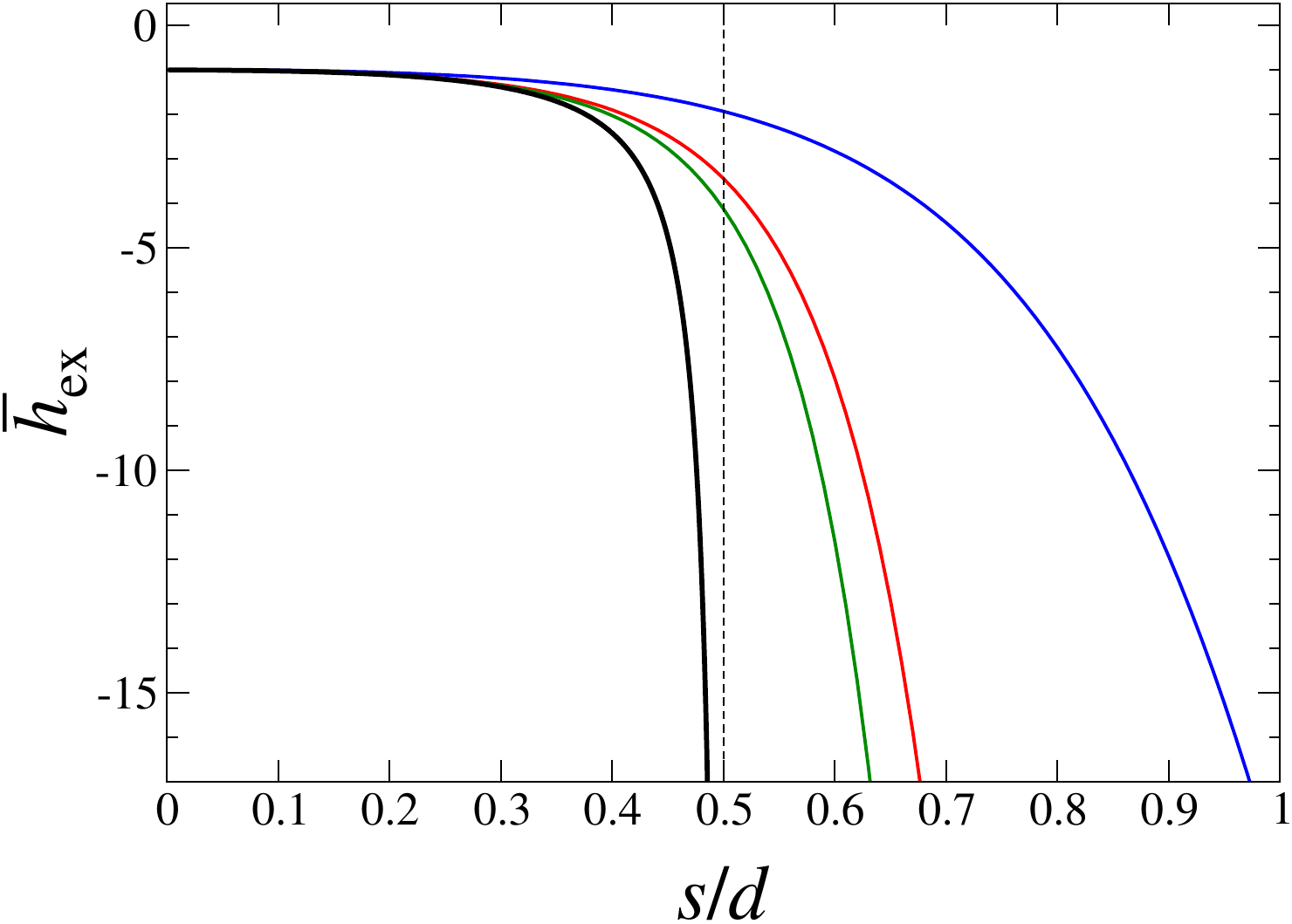}}
 \caption{Upper bound on the entropy cost per particle due to density correlations, as a function of $s/d$.  The entropy is rescaled by the prefactor given in equation~(\ref{divergence}). The lowest, black curve shows the function $\psi(s/d)$, giving the rescaled entropy bound for $s<d/2$ and diverging as $s/d\to 1/2$; see equation~(\ref{divergence}). From top to bottom, the blue, red, and green curves give the rescaled entropy bound for $d=2$ and the upper cutoff $\lambda K=10,50,100$, respectively, according to equation~(\ref{hex_cutoff}).} 
 \label{fig:psi}
\end{figure}

However, the situation is more subtle. The function $\psi(x)$ is shown by the lowest, black curve in figure~\ref{fig:psi}. It tends to $-\infty$ as $x\rightarrow 1/2$, i.e., as $s$ tends to $d/2$ from below. For $s\geq d/2$, $\hup\rightarrow -\infty$. Since $\hup$ is an upper bound, this result implies that the entropy itself diverges.

The divergence indicates a failure of the high-temperature theory presented above. It comes from the large-wavevector limit of the integration in equation~(\ref{hex}), i.e., not from the long range of the interaction but rather from the strong repulsion at small $r$. The larger the value of $s$, the stronger the interaction of nearby particles {which, according to the theory above, would result in strongly correlated
large-$k$ density modes. If the approach of $S(\veck)$ to $1$ at large $k$ is not sufficiently fast, the negative contribution of these strong correlations to the entropy diverges.
%becomes dominant. 
Introducing an upper cutoff $K$ for $|\veck|$ to regularize this divergence, we readily find
\begin{equation}
    \hup \sim 
    \left\{ \begin{array}{ll}
    -K^{2s-d}, & d/2 < s < d \\
    -\ln K,\ \ \ & s=d/2.
    %\\
    %-K^d, & s = d \\
    %-K^d \ln K & s >d.
    \end{array} \right.
\end{equation}
In more detail, the integration in equation~(\ref{hex}) can be performed with the upper cutoff $K$, yielding
\begin{eqnarray}
\label{hex_cutoff}
  \hspace{-.7cm}
   \hup =&& \frac{{\cal S}_d} {2(2\pi)^d d \, \rhob}\, K^d \left[ \ln S(K) + (1-\alpha/d) \, _2F_1 \left( 1, d/\alpha, 1+d/\alpha, -(\lambda K)^\alpha \right) \right]
   \\
   && %\xrightarrow{\lambda K\gg 1} 
   \overset{\lambda K\gg 1}{\simeq} 
   -\frac{{\cal S}_d} {4(2\pi)^d\, \rhob \lambda^d} \times
   \left\{ \begin{array}{ll}
   \frac{1}{2s-d} (\lambda K)^{2s-d},\ \ \ & d/2< s< d \\
   \ln(\lambda K), %- 1/d 
   & s=d/2, \end{array} \right.
\label{hex_cutoff_highT}
\end{eqnarray}
where $_2F_1$ is the hypergeometric function. The results of equation~(\ref{hex_cutoff}) for $d=2$ and $\lambda K=10,50,100$ are shown by the upper three curves in figure~\ref{fig:psi}. As the cutoff is increased, the curves approach the divergent curve for infinite $K$.

The upper-$k$ cutoff must affect various measurable quantities derived from the entropy. For example, assuming that the bound is tight, $\hex\simeq\hup$, as expected at sufficiently high temperature, we obtain the excess heat capacity and mean square fluctuation of the potential energy per particle as
\begin{equation}
    c_{\rm ex} = T \frac{\partial \hup}{\partial T}, %= -\frac{d}{d-s} \hup,
    \ \ \ \ 
    \langle (\Delta u)^2\rangle = T^2\,c_{\rm ex}.
\label{cex}
\end{equation}
For $s<d/2$ we get from equations~(\ref{divergence}) and (\ref{cex}) $\langle (\Delta u)^2\rangle \sim T^2 \lambda^{-d} \sim T^{(d-2s)/(d-s)}$. (The anomaly for $s\in(d/2,d)$ is seen also here; for these values of $s$ the energy fluctuations supposedly decrease rather than increase with temperature.)

The particles being point-like, a natural choice for the lower cutoff length is $\ell$, the distance below which the pair-repulsion $v_0/r^s$ is stronger than $T$,
\begin{equation}
    \ell = (v_0/T)^{1/s},\ \ \ \ K = c/\ell,
    \label{ell}
\end{equation}
where $c$ is an unknown numerical factor. Substituting this cutoff in equation~(\ref{hex_cutoff_highT}) and using equation~(\ref{cex}), we find
\begin{equation}
\label{Du_high}
    \mbox{high temperature:}\ \ \ \langle (\Delta u)^2\rangle \sim \left\{
    \begin{array}{ll}
    \rhob^{\, s/(d-s)}\, T^{(d-2s)/(d-s)},\ \ \ \ & s<d/2 \\
    \rhob\, T^{(2s-d)/s}, & d/2<s<d \\
    \rhob\, \ln (T/\rhob), & s=d/2.
    \end{array} \right.
\end{equation}
The prefactors can readily be worked out from the equations above. Thus, the dependencies of the potential energy fluctuations on temperature and density are qualitatively different for $s<d/2$ and $s\geq d/2$. Equation~(\ref{Du_high}) holds at sufficiently high temperature because it assumes that the cutoff $K$ is sufficiently large, $\lambda K~\sim~T^{(d/s)/(d-s)}~\gg~1$.

As $\ell$ does not represent a sharp barrier, $K$ cannot be a sharp cutoff. For $s\geq d/2$ we expect the amplitudes of density modes with $k\gtrsim K$ to be finite but become less correlated than predicted by equation~(\ref{Sk}).
Namely, $1-S(k\gtrsim K)$ should decay faster than $(\lambda k)^{-\alpha}$. As the temperature is increased, $K\sim T^{1/s}$ increases, and the deviation of $S(\veck)$ from equation~(\ref{Sk}) should occur at larger and larger wavevectors. At a sufficiently high temperature the cutoff will have to be replaced by a finite particle size, which is not included in the present theory.

As the temperature decreases, $\ell$ increases and the cutoff $K$ decreases. If we conjecture that equation~(\ref{hex_cutoff}) should qualitatively hold also in the low-temperature (equivalently, high-density) regime, $\lambda K\ll 1$, we obtain $\langle (\Delta u)^2\rangle \sim \rhob^{-1} T^{d/s+2} \ln T$. This dependence on $T$ is stronger than the high-temperature one given in equation~(\ref{Du_high}), for both $s<d/2$ and $s\geq d/2$.  

%As mentioned in Sec.~\ref{sec:scaling}, by modifying the structure factor at small $k$ to make it hyperuniform, the system  avoids small-$k$ modes with a divergent potential energy. Similarly, for $s\geq d/2$, one expects the structure factor to deviate at large $k$ from equation~(\ref{Sqvariational}) such that the divergent energy fluctuations are avoided. Equation~(\ref{Sqvariational}) gives, at large $k$, $1-S(\veck)\sim (\lambda k)^{-\alpha}$, with $\alpha=d-s \in [0,d/2]$ for the problematic range $s\in [d/2,d]$. The divergence of the entropy and energy fluctuations will be avoided if, instead, $1-S(\veck)\sim (\lambda k)^{-\eta}$ at large $k$, with $\eta > d/2$. 

\section{Simulations}
\label{sec:simulations}
We run swap Monte Carlo simulations of a two-dimensional system with $N$ point-like particles, interacting via the truncated and shifted version of the pair-potential $v(\vecr)$ of equation~(\ref{vr}):
\begin{equation}
    v_c(\vecr) = 
    \begin{cases}
    \dfrac{v_0}{r^s} - \dfrac{v_0}{r_c^s} & \mathrm{if} \,\,\, r \leq r_c, \\
    0 & \mathrm{else,}
    \end{cases}
\label{vrcut}
\end{equation}
with an interaction-cutoff distance $r_c$ of the order of the system size. The introduction of $r_c$ has three motivations. (a) By exploring the effect of $r_c$ on the results we address the  complications associated with finite-size (boundary) effects in systems with long-range interactions~\cite{Ewald1921,Mazars2011}. (b) Sensitivity of the results to the value of $r_c$ indicates dominance of large-wavelength fluctuations. (c) Long-ranged interactions without a cutoff require the inclusion of a one-body potential (i.e., a neutralizing background as in the case of the one-component plasma model) to ensure  stability \cite{Hansen1973,Sal2024JCP}. %whereas the volume integral with cutoff converges... so no need for one-body potential}.  
In the remainder of this section we derive the equations and present the results for $d=2$. 

We run $250$ independent $NVT$ simulations at each prescribed temperature by starting from different initial configurations. Each simulation runs for $2 \times 10^5$ Monte Carlo sweeps and is sampled every $5\times10^2$ sweeps where a sweep corresponds to an attempted move of each of the $N=2000$ particles under periodic boundary conditions in a square box of side length $L$. The first $1 \times 10^5$ sweeps contain swaps between two randomly selected individual particles with a swap attempt probability $p=0.3$ \cite{Parisi2001,Coslovich2017}, followed by further $1 \times 10^5$ sweeps with standard Metropolis algorithm without swapping. Our results are sampled from the latter part without swapping and averaged over the $250$ runs. The step size for attempted moves is adjusted such that the acceptance rate is approximately 40\%. 

The simulations are performed for a low-density fluid with $\rhob=0.002$ and at temperatures  between $T/v_0=0.2$ and $T/v_0=200$, using a unit of length equal to $L/1000$. We study interaction potentials with $s=0.5, 1.2, 1.5$ and  $r_c=L/2,L/8$. Although we do not investigate the low-temperature behavior in detail in the present work, we observed the formation of a hexagonal crystal in our finite systems upon  quenching down to absolute zero in a few test cases. In the studied temperature range we do not observe crystallization. Thus, the systems considered have a crystalline ground state with a melting temperature lower than the lowest temperature studied here. 

To compare our numerical results with the theory, we first provide an analytical expression for the structure factor with the interaction-cutoff distance $r_c$. The 2D Fourier transform of the truncated potential (\ref{vrcut}) for $s<2$ reads as
\begin{equation}
\tilv_c(\veck) = 2\pi r_c^{2-s}v_0 \left( - \dfrac{J_1(kr_c)}{k r_c} - \dfrac{{_1}F_2(1-s/2, 1,2-s/2, -(kr_c)^2/4)}{s-2} \right),
\label{vkcut}
\end{equation}
where $J_1$ is the Bessel function of the first kind, and $_1F_2$ denotes the hypergeometric function. 
Specializing the structure factor of equation~(\ref{Sk}) to the modified potential of equation~(\ref{vkcut}) yields 
\begin{equation}
 S_c(\veck) = \frac{r_c^s}{r_c^s - 2\pi r_c^{2} (v_0/T) \rhob B(k)},
\label{Skcut}
\end{equation}
where $B(k)$ is given as 
\begin{equation}
 B(k) =  \dfrac{J_1(k r_c)}{k r_c} + \frac{{_1}F_2(1-s/2, 1,2-s/2, -(kr_c)^2/4)}{s-2},
\label{Bk}
\end{equation}
which converges to $B(k \to 0) = 1/(s-2) + 1/2$ in the limit of infinite wavelength. Consequently, the limit of $S_c(\veck)$ in equation~(\ref{Skcut}) becomes
\begin{equation}
 S_c(k \to 0) =  \frac{1}{1 + \dfrac{\pi r_c^{2-s} s (v_0/T) \rhob}{2-s}} .
\label{Sklim}
\end{equation}
As $r_c \to \infty$ we recover the hyperuniformity, $S(k \to 0) \to 0$, for pure power-law potentials with $s<d$. 

\begin{figure}[ht!]
 \centerline{ \includegraphics[width=0.45\textwidth]{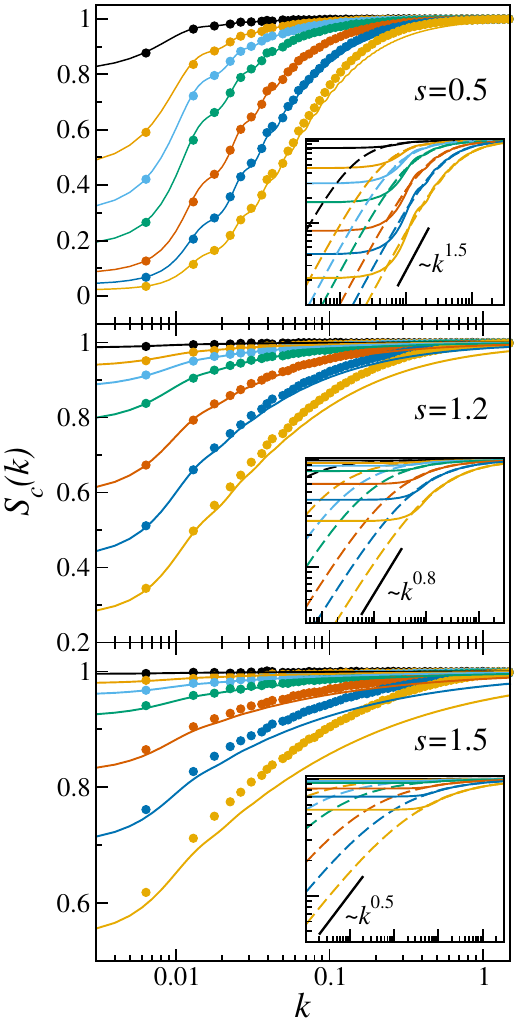} }
  \caption{Structure factors obtained from MC simulations (circles) and from the variational analysis (equation~(\ref{Skcut}), solid lines). The different data sets, from top to bottom, correspond to decreasing temperature, $T/v_0=100,20,10,5,2,1,0.5$. Upper, middle, and lower panels show the results for the interaction exponents $s=0.5, 1.2, 1.5$, respectively (i.e., $s$ both smaller and larger than $d/2$). The interaction cutoff is $r_c=L/2$. The unit of length is $L/1000$. The insets compare the theoretically predicted structure factors with interaction cutoff $r_c=L/2$, $S_c(k)$ (equation~(\ref{Skcut}), full lines), with those for pure power-law repulsion without a cutoff, $S(k)$ (equation (\ref{Sqvariational}), dashed lines). The thick black lines in the insets indicate the scaling behavior of $S(k)$ in the small-$k$ limit, $S(k) \sim k^{d-s}$ with $d=2$.}
 \label{fig:sofks}
\end{figure}

In figure~\ref{fig:sofks} we show $S_c(k)$ as obtained from MC simulations (circles), together with the theoretical prediction given in equation~(\ref{Skcut}) (full lines), for $s=0.5,1.2,1.5$ (i.e., $s$ both smaller and larger than  $d/2=1$), using a cutoff distance $r_c=L/2$. In all panels the temperatures are in decreasing order from top to bottom, $T/v_0=100,20,10,5,2,1,0.5$. The insets demonstrate the effect of the interaction cutoff as obtained theoretically in equations~(\ref{Sqvariational}) and (\ref{Skcut}). The structure factors $S(k)$ in the absence of a cutoff (dashed lines) decay to zero for $k\to 0$ as $k^{2-s}$. The structure factors $S_c(k)$ with a cutoff (solid lines) level off at sufficiently small $k$, reaching their limit value $S_c(k\to 0)$ given in equation~(\ref{Sklim}).

\begin{figure}[ht!]

  \centerline{  \includegraphics[width=0.45\textwidth]{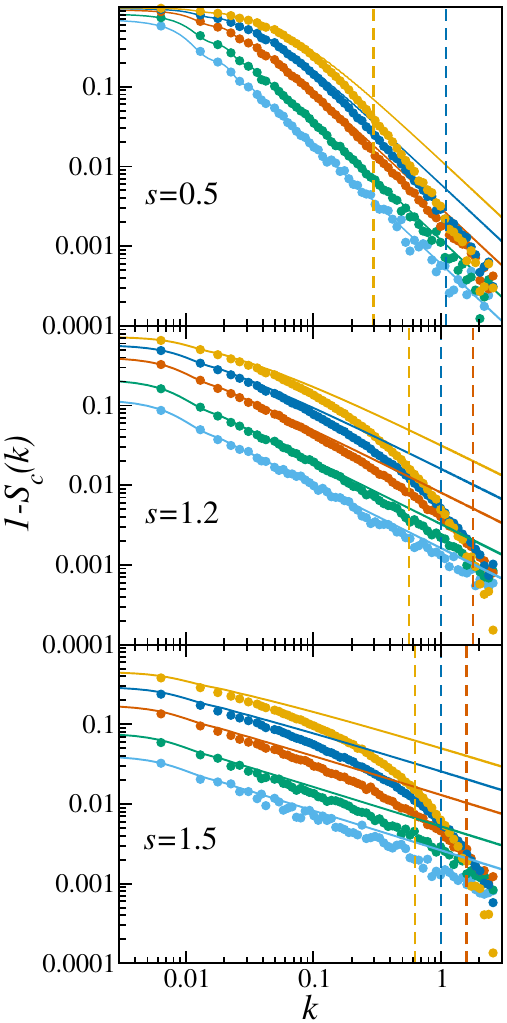} }
  \caption{Approach of the structure factor to 1, $1-S_c(k)$, from MC simulations (circles) and theory (equation~(\ref{Skcut}), solid lines). The three panels show results for $s=0.5, 1.2, 1.5$. The temperature are, from top to bottom, $T/v_0 = 0.5,1,2,5,10$, with color coding as in figure~\ref{fig:sofks}. The vertical dashed lines indicate the wavevectors $k=1/\ell$ (equation~\ref{ell}) for $T/v_0=0.5,2,5$ (left to right).}
  %, above which $S_c(k)$ is expected to approach to $1$ faster for $s\ge d/2$ than for $s<d/2$.}
 \label{fig:sofks_2}
\end{figure}

As seen in figure~\ref{fig:sofks}, which is focused on small $k$, the simulation results and theoretical predictions agree remarkably well for small $s$ and high temperature, including fine oscillatory features in $S_c(k)$ (upper panel; see more below). For larger $s$ and smaller $T$, the deviation of the theory from the simulation is larger and begins at a smaller value of $k$. This is probably due to higher-order correlations, neglected by the theory, which should become less significant with increasing temperature and increasing range of interaction.
%\eco{I guess I am tryin to argue that there is another crossover other than 1/l here.}
Still, the simple high-temperature theory, which considers only two-point correlations, captures well the behavior at  sufficiently small $k$ even for the lower temperatures. 
%And for a given $T$, the mean-field regime sets in at ever smaller $k$ upon increasing $s$. The strong inter-particle repulsion at small scales for $s>d/2$ therefore requires larger-scale fluctuations that can 'shield' the effects of higher-order correlations than for $s<d/2$. 

The structure factor $S_c(k)$ exhibits small oscillations (see the upper panel in figure~\ref{fig:sofks}). These oscillations are a result of the interaction cutoff distance $r_c$. They are absent in $S(k)$ for pure power-law potentials. The periodicity $\Delta k$ of the oscillations corresponds to $2\pi/\Delta k \approx r_c$. As most studies on fluids focus on either short-ranged interactions or purely long-ranged interactions without a cutoff (using, e.g., Ewald sums), this oscillatory effect, to our best knowledge, has not been presented before. The fact that the oscillations are pronounced for $s=0.5$ (upper panel) and negligible for $s=1.2,1.5$ (lower two panels) indicates the sensitivity of the longer-range interactions (small $s$) to the interaction cutoff.

In figure~\ref{fig:sofks_2} we turn our attention to the large-$k$ modes. To examine how the structure factor approaches $1$ at large $k$ we plot $1-S_c(\veck)$ as obtained from MC simulations and from equation~(\ref{Skcut}). 
%
%(full lines) for $s=0.5,1.2,1.5$ in the upper, middle, and lower panel, respectively. We use the same color code as in figure~\ref{fig:sofks}, and for the sake of clarity, we present here fewer temperatures (from top to bottom: $T/v_0 = 0.5,2,5,10$). The dashed vertical lines correspond to $k=1/l$ for $T/v_0=0.5,2,5$, cf.\ equation~(\ref{ell}).
%
%The small-scale fluctuations play a pivotal role in regularizing the entropy. 
As predicted in section~\ref{sec:fluctuations}, density fluctuations on this small scale (large $k$) are 
%strongly suppressed. 
enhanced, decorrelating the density modes at these length scales. Thus the fluctuations of particle number become more quickly Poissonian with increasing $k$ than what would be expected from the strong inter-particle repulsion.
This is manifested in large deviations of $1-S_c(k\gtrsim K)$ from the predictions of the high-temperature theory, equations~(\ref{Sk}) and (\ref{Skcut}). The deviations become stronger and sharper as $s$ is increased and as the temperature is decreased, as anticipated. Figure~\ref{fig:sofks_2} shows also the locations of the crossover wavevectors, $k\sim 1/\ell$, beyond which the 
decorrelation should occur (dashed vertical lines). The $k$ values where the structure factor starts deviating from the theory are correlated with $1/\ell$, as conjectured; yet, for $s=1.2, 1.5$ they are significantly smaller than $1/\ell$. This wider range of enhanced
%suppressed 
fluctuations expands the validity range of our main results as it implies that the enhancement
%suppression 
occurs already for wavelengths much larger than the typical molecular size. The deviations for $s=1.5$ are sharper than for $s=1.2$ without an appreciable change in $1/\ell$, highlighting the effect of increasing $s$ on the magnitude of enhancement
%suppression 
in addition to its location. Deviations from the theory are present also for $s=0.5<d/2$, which has not been anticipated. They are much smaller and less sharp than for $s>d/2$ and begin at $k$ values closer to $1/\ell$ (upper panel). Consequently, they do not have a significant effect on the thermodynamics, as we show next.
%Moreover, figure~\ref{fig:sofks_2} seems to indicate that $1-S_c$ decays faster for $s=1.5$ and $s=1.2$ than for $s=0.5$ at sufficiently large $k$, suggesting strong suppression of large-$k$ modes for $s>d/2$ in addition to small-$k$-mode suppression for both $s<d/2$ and $s>d/2$.  

\begin{figure}[ht!]
  \centerline{\includegraphics[width=0.5\textwidth]{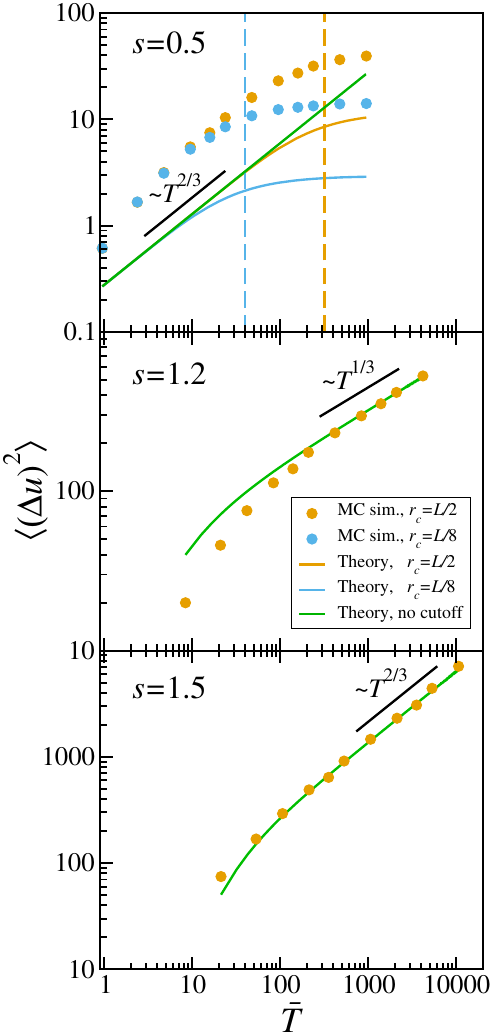} }
  \caption{Fluctuations of the potential energy per particle $\langle (\Delta u)^2\rangle$ as a function of rescaled temperature $\bar{T}= T \rhob^{\,-s/2} /v_0$ for $s=0.5, 1.2, 1.5$, as obtained from MC simulations (circles) and analytical theory (solid lines). The upper panel shows results for two values of the interaction cutoff distance, $r_c=L/2, L/8$, and the analytical result in the absence of cutoff. %For $s=0.5<d/2$ the energy fluctuations are sensitive to the value of $r_c$ at large $\bar{T}$. 
  The vertical dashed lines indicate the temperature above which the screening length $\lambda$ becomes larger than $r_c$.  In the lower two panels for $s=1.2, 1.5>d/2$ the solid curves show analytical results using the fitting parameters $c=15, 1.5$, respectively. 
  %The theoretical curves for different $r_c$ are essentially indistinguishable (not shown). 
  Black lines indicate the power laws predicted in equation~(\ref{Du_high}). The fluctuation axis is rescaled by $\rhob^{\,-s} /v_0^2$.}
 \label{fig:nrgflucs}
\end{figure}

In section~\ref{sec:fluctuations} we have predicted a qualitatively different dependence of the potential energy fluctuations on temperature for $s<d/2$ and $s\geq d/2$; see equation~(\ref{Du_high}). 
In~figure~\ref{fig:nrgflucs} we show numerical and analytical results for the fluctuations of the potential energy per particle  $\langle (\Delta u)^2\rangle$ as a function of $T$ for $s=0.5,1.2,1.5$. 
The analytical results are obtained using equations~(\ref{hex_cutoff}), (\ref{cex}), (\ref{ell}), and (\ref{Skcut}).

Comparing the uppermost panel of figure~\ref{fig:nrgflucs} with the lower two, we clearly see the qualitative differences between the two categories. First, the data follow the different power laws predicted at high temperature (equation~(\ref{Du_high})), $(d-2s)/(d-s)=2/3$ for $s=0.5$, and $(2s-d)/s=1/3, 2/3$, for $s=1.2, 1.5$, respectively.
In addition, we find a very different effect of the interaction cutoff distance $r_c$ on the energy fluctuations. For $s=0.5$ the cutoff distance makes the fluctuations saturate to a constant value at high temperature. (Recall that the  fluctuations are of the potential energy only.) The constant value increases with $r_c$, as it should assuming that the saturation disappears in the limit of infinite $r_c$. The crossover between the power law and the saturation occurs around the temperature for which $\lambda(T)=r_c$ (see the dashed vertical lines). Above that temperature the screening length exceeds the interaction cutoff distance, leaving no room for long-range effects such as hyperuniformity. In other words, when $\lambda(T)>r_c$ the interaction is essentially short-ranged. Thus the sensitivity to $r_c$ indicates the dominant contribution of large-wavelength fluctuations for $s=0.5<d/2$. 
By contrast, the fluctuations for $s=1.2, 1.5>d/2$ as obtained from theoretical calculations are essentially independent of $r_c$, implying the dominance of small-wavelength fluctuations. For these two values of $s>d/2$ the simulation data with interaction cutoff show good agreement with the theory without interaction cutoff at relatively high temperatures, where the theory should hold, once we fit the free parameter $c$ in the relation $K=c/\ell$ (equation~(\ref{ell})). The fitted values, however, differ substantially between the two cases ($c=15, 1.5$ for $s=1.2, 1.5$, respectively). 
%we have used $c=15$ and $c=1.5$ to achieve a reasonable quantitative agreement with the simulation data for $s=1.2$ and $s=1.5$, resp.    

%On the other hand, for  $s<d/2$, the entropy and hence the energy fluctuations are dominated by strongly correlated large-scale density modes that renders the system sensitive to the cutoff distance in the pair-potential. In the upper panel of figure~\ref{fig:nrgflucs}, we examine the effect of the cutoff distance $r_c$ on energy fluctuations by investigating the two systems with $r_c=L/2$ and $r_c=L/8$. We find that the fluctuations saturate at high temperatures (orange and blue solid lines) as opposed to increasing fluctuations without a cutoff for the same $s$ (green solid line). The crossover to the saturation region is marked by the dashed vertical orange and blue lines in the upper panel of figure~\ref{fig:nrgflucs}. These lines denote the temperature at which the screening length $\lambda$ equals $r_c$ for the two cutoff distances $r_c=L/2$ and $r_c=L/8$. \eco{Can you help me please. I need to say why we choose lambda=rc, but can't make the connection. So, we have hyperuniformity beyond lambda, and it's limited by rc. If $\lambda \ge rc$, we loose the hyperuniformity. Fine. But can I see explicitly somewhere somehow that energy flucs saturate beyond rc? I tried to use your scaling argument from Sec. 3, inserting the truncated v(r) into $\langle |\delta u| \rangle \sim \langle (\delta n)^2\rangle\, v(R)$, bu couldn't get to anywhere..}

Overall, the simulations confirm the main predictions of the high-temperature theory, namely, the power laws, general shapes of the curves, and sensitivity to $r_c$. However, the large quantitative discrepancy for $s=0.5$ (uppermost panel in figure~\ref{fig:nrgflucs}), as well as the very different values of the free parameter $c$ required to fit the curves for $s=1.2$ and $s=1.5$ (lower two panels), show that the agreement between simulation and theory is still far from being quantitative.

%\ell=(v_0/ (T + v_0/r_c))^{1/s}$

\section{Discussion}
\label{sec:discussion}

Our analytical and numerical results highlight the distinction between repulsive Riesz gases with interaction decay power $s<d/2$, exhibiting 
%large-scale
hyperuniformity, and ones with $d/2 \leq s <d$, which exhibit hyperuniformity as well as small-scale decorrelation. The entropy of the former is dominated by the cost of suppressed large-scale fluctuations, equation~(\ref{divergence}). The latter's entropy is dictated by the contribution of 
small-scale correlations, equations~(\ref{hex_cutoff_highT}) and (\ref{Du_high}). We have reached these conclusions by noticing the failure of a simple high-temperature theory to yield a physical entropy for the second category. We showed that the theory is regularized by an intrinsic cutoff distance $\ell$, akin to the electrostatic Bjerrum length. Our results are practically significant provided that $\ell$ is larger than the physical lower cutoff distance set by the molecular size. This is the typical case (see examples below). Moreover, as seen in figure~\ref{fig:sofks_2}, the relevant effects set in already at $k$ values much smaller than $\ell^{-1}$, which further broadens the range of applicability of our results. At sufficiently high temperature, the regularization of the theory will come from the molecular size $a$ rather than the intrinsic $\ell$, changing the dependence of entropy on density and temperature. The modified expressions are obtained from equations~(\ref{hex_cutoff}) and (\ref{hex_cutoff_highT}) by changing $\lambda K$ to $\lambda/a$.

The simple theory gives excellent results for $S(\veck)$ at high temperature for almost all $k$ values (figure~\ref{fig:sofks}). However, it fails at sufficiently high $k$ for {\em all} temperatures (figure~\ref{fig:sofks_2}). For $s<d/2$ the consequences are unimportant, but for $s\geq d/2$ they are far-reaching.
More accurate theories than the simple high-temperature starting point used here were developed for systems with long-range interactions \cite{CampaPhysRep2009,BouchetPhysA2010,LewinJMathPhys2022}. As they did not focus on short-wavelength fluctuations and their contribution to the entropy, they seem to have missed the distinction between the categories of $s<d/2$ and $s\geq d/2$. Such rigorous analyses are clearly warranted, for example, to obtain the crossover occurring around $k\sim\ell^{-1}$ as an analytical result rather than a numerically confirmed Ansatz. This will also make the related free parameter $c$ unnecessary. Better quantitative agreement with simulation results for the energy fluctuations is also desirable. A possible route for such a future extension is to derive a more accurate entropy bound than equation~(\ref{hex}), taking into account higher-order correlations.

Notwithstanding, the high-temperature theory seems to become accurate for all temperatures at sufficiently small $k$, and the agreement with the simulations improves with decreasing $s$ (see figure~\ref{fig:sofks}). This success is related to the fact that the theory correctly predicts the hyperuniformity power law at small $k$, $S(k)\sim k^\alpha$, as obtained from the general scaling analysis (section~\ref{sec:scaling}) and rigorous sum rules~\cite{BlumPRL1982,MartinRMP1988}. Thus, a similar analysis should be useful for studying large-scale fluctuations (i.e., hyperuniformity) also at low temperature, as long as the system's symmetry is not broken, for example, upon approaching a glass transition. These issues will be the subject of a forthcoming publication.
%(\eco{the following paragraph is written only for us to make the connection to glass paper. Not sure if we should mention it here.})   
%We conjecture that the agreement between simulations and theory at sufficiently small wavevectors continues to exist for even much lower temperatures as long as the symmetry of the system is not broken. Sufficiently large-scale density fluctuations can be described well by pair correlations only. Assume a system with long-ranged repulsion that undergoes a glass transition: the amorphous solid can be characterized by mean-field theory so long we consider sufficiently large-wavelength modes(???).

An example of a system which should exhibit the behavior studied here is a dilute cloud of charged particles near a conducting surface. Due to image charges the repulsive pair-potential has $d=3$ and $s=2>d/2$. (Only if two ions happen to be at the same distance from the surface do they interact via a potential with $s=3$.) For ions of elementary charge $e$ at a typical distance $h\sim 1$~nm from the surface, at room temperature in vacuum or a gas phase, we get $\ell \sim (e^2 h/T)^{1/2} \sim 1$--$10$~nm, larger than a typical ion size. Thus, the energy fluctuations of the ion cloud should be dominated by wavelengths of order $\ell$ and smaller, and obey the scaling given in equation~(\ref{Du_high}), $\langle (\Delta u)^2 \rangle \sim \rhob T^{1/2}$.

The enhancement
%suppression 
of small-scale density fluctuations should hold also for $s>d$, where 
%large-scale 
hyperuniformity does not occur. An experimentally relevant example is a fluid monolayer of dipolar particles aligned perpendicular to a liquid-gas interface. The dipoles interact via a repulsive power-law potential with $d=2$ and $s=3$. The interaction is dominated by the gas phase, whose dielectric constant is much smaller than the liquid's. For molecules with a dipole moment $p \sim 1$~D, at room temperature, we get $\ell \sim (p^2/T)^{1/3} \sim 1$~nm, larger than the typical size of a small dipolar molecule. According to equation~(\ref{Du_high}), the energy fluctuations should scale as $\langle (\Delta u)^2 \rangle \sim \rhob T^{4/3}$. However, the larger the interaction decay power $s$, the smaller the cutoff distance $\ell$. For example, in the case of the Lennard-Jones potential with the conventionally defined parameters $\epsilon$ and $\sigma$, we have $\ell \sim (\epsilon/T)^{1/12} \sigma$. No matter how large the repulsion amplitude $\epsilon$ may be, $\ell$ is invariably comparable to the molecular size $\sigma$, making the results of the present study irrelevant in this case. Indeed, power-law potentials of large $s$ were found to behave similarly to Gaussian potentials despite their very different small-$r$ dependence~\cite{PrestipinoJCP2005}. 

To circumvent the issue of system-size effects, and to further highlight the distinction between the two categories, we have used in the simulations a potential that vanishes beyond a cutoff distance $r_c$ of the order of the system size. For $s<d/2$, changing the value of $r_c$ affects the density and energy fluctuations, and the effect is qualitatively captured by the high-temperature theory (figure~\ref{fig:nrgflucs}, uppermost panel). By contrast, the systems with $s\geq d/2$, which are dominated by small-scale fluctuations,  are found to be insensitive to the cutoff.
Also noteworthy is the effect of $r_c$ on the structure factor $S_c(k)$ (figure~\ref{fig:sofks}). For $s<d/2$ the cutoff introduces pronounced oscillations with periodicity $\sim 1/r_c$, whereas for $s>d/2$ the oscillations are negligibly small. In all cases the interaction cutoff makes the small-$k$ dependence of $S_c(k)$ saturate to a constant, removing the 
%large-scale 
hyperuniformity. As $r_c\rightarrow\infty$ the oscillations disappear and the small-$k$ dependence becomes a power law, restoring 
%large-scale 
hyperuniformity. 

The distinctive behaviors of systems governed by long-range repulsion with $d/2 \leq s < d$ must affect their dynamics. This calls for separate studies.

\ack

We thank J\"urgen Horbach, Yitzhak Rabin, and Salvatore Torquato for illuminating discussions. This research has been supported by a joint grant from the Israel Science Foundation and the National Natural Science Foundation of China (ISF-NSFC Grant No.\ 3159/23). E.C.O further acknowledges support from RFIS-II Grant by the National Natural Science Foundation of China (Grant No. 12350610238). 

~~\\
~~\\

\bibliographystyle{iopart-num}

\bibliography{jpcm24}

\end{document}